# Chemical doping and high pressure studies of layered $\beta$-PdBi$_2$ single crystals

Kui Zhao[1], Bing Lv[1], Yu-Yi Xue[1], Xi-Yu Zhu[1,2],
Liangzi Deng[1], Zheng Wu[1] and C. W. Chu[1,3]

[1] Texas Center for Superconductivity and Department of Physics, University of Houston, Houston, TX 77204-5002.
[2] Department of Physics, Nanjing University, Nanjing, China.
[3] Lawrence Berkeley National Laboratory, Berkeley CA 94720

**Abstract:**

We have systematically grown large single crystals of layered compound $\beta$-PdBi$_2$, both the hole-doped PdBi$_{2-x}$Pb$_x$ and the electron-doped Na$_x$PdBi$_2$, and studied their magnetic and transport properties. Hall-effect measurement on PdBi$_2$, PdBi$_{1.8}$Pb$_{0.2}$, and Na$_{0.057}$PdBi$_2$ shows that the charge transport is dominated by electrons in all of the samples. The electron concentration is substantially reduced upon Pb-doping in PdBi$_{2-x}$Pb$_x$ and increased upon Na-intercalation in Na$_x$PdBi$_2$, indicating the effective hole-doping by Pb and electron-doping by Na. We observed a monotonic decrease of superconducting transition temperature ($T_c$) from 5.4K in undoped PdBi$_2$ to less than 2K for x > 0.35 in hole-doped PdBi$_{2-x}$Pb$_x$. Meanwhile, a rapid decrease of $T_c$ with the Na intercalation is also observed in the electron-doped Na$_x$PdBi$_2$, which is in disagreement with the theoretical expectation. In addition, both the magnetoresistance and Hall resistance further reveal evidence for a possible spin density wave (SDW)-like transition below 50K in the Na-intercalated PdBi$_2$ sample. The complete phase diagram is thus established from hole-doping to electron-doping. Meanwhile, high pressure study of the undoped PdBi$_2$ shows that the $T_c$ is linearly suppressed under pressure with a $dT_c/dP$ coefficient of -0.28K/GPa.

**Introduction:**

Low dimensional compounds, with simple structure building motifs and weak bonding force between layers, have generated much research interest within condensed matter physics over the past several decades. These low dimensional materials have displayed a variety of unusual physical phenomena such as charge density waves in transition metal chalcogenides[1-2]; spin density waves in parent Fe-pnictide superconductors[3-4]; topological order in Bi$_2$Se$_3$[5-6]; and superconductivity in many compounds such as MgB$_2$[7], doped ZrNCl[8], etc. In the binary Pd-Bi alloy family, several phases with different structures have been studied[9] in the past, in which the PdBi$_2$ is found to crystallize in two different layered structures with a low temperature α-phase below 380°C and a high temperature β-phase between 380°C and 490°C[10]. The α-PdBi$_2$ crystalizes in a layered monoclinic (c2/m) structure with a six-coordinated PdBi$_6$ building motif, while the β-PdBi$_2$ forms a layered tetragonal CuZr$_2$-type structure (I4/mmm) with an eight-coordinated PdBi$_8$ building motif. Every four Bi-atoms in the β-PdBi$_2$ are face-shared with the neighboring CsCl-type PdBi$_8$ motif and therefore form a PdBi$_{8/4}$=PdBi$_2$ layer. The resulting PdBi$_2$ layers are packed alternately and form the body centered tetragonal structure, as shown in the inset of Fig. 1(a). The interlayer spacing between the alternating PdBi$_2$ layers is rather large, with interlayer Bi-Bi distance of ~3.8Å, indicating that there is no effective bonding between those layers. Various phases in the Pb-Bi system have been identified as superconductors[11], such as α-PdBi with $T_c$~3.8K, α-PdBi$_2$ with $T_c$ ~1.73K, β-PdBi$_2$ with $T_c$ ~ 4.25K, and Pd$_{2.5}$Bi$_{1.5}$ with $T_c$ ~3.7-4K. Early studies showed that β-PdBi$_2$ had the highest $T_c$ among them, 4.25 K [11], and it was recently shown that the $T_c$ could be further raised to 5.4K by improving the sample quality[12]. However, neither details of chemical doping nor high pressure effect have been studied previously in β-PdBi$_2$. The prior specific heat study on β-PdBi$_2$ suggests that it is a multi-band/multi-gap superconductor[12], which agrees with the results from first principle calculations[13]. Theoretical calculation shows that the density of states (DOS) around the Fermi level is dominated by the Pd 4$d$ and Bi 6$p$ states, and that the Fermi level is located on a positive slope below a DOS peak. Therefore, hole doping is expected to shift the Fermi level away from the DOS peak, resulting in a decrease of the DOS at the Fermi level while electron doping will increase it. Therefore, one would expect a decrease of $T_c$ upon hole doping and an increase of $T_c$ under electron doping. Under this motivation, we decided to carry out systematic hole doping (substitution of Bi with Pb) and electron doping (Na-intercalation) studies on the β-PdBi$_2$ system. However, we found that the $T_c$ was suppressed in both cases, the reasons for which will be discussed in this paper. Meanwhile, the high pressure study was also carried out on the β-PdBi$_2$ single crystal and we observed a suppression of $T_c$ upon applied pressure with a $dT_c/dP$ coefficient of -0.28K/GPa. The phase diagrams, both upon chemical doping and under high pressure, are presented.

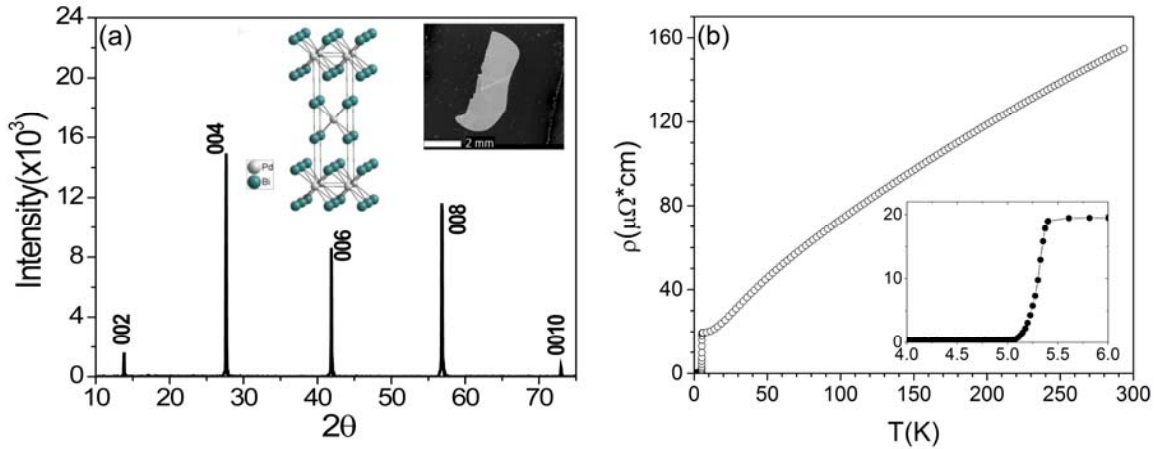

Fig.1 (a) XRD pattern of β-PdBi$_2$ with preferred orientation along c axis; the inset shows the crystal structure and one representative crystal SEM image. (b) Resistivity of β-PdBi$_2$ from 2K to 300K; the inset displays resistivity data between 4K and 6 K.

**Experimental details:**

The β-PdBi$_2$ single crystals were synthesized through a melt-growth method. Stoichiometric amounts of Pd and Bi grains were sealed in an evacuated quartz tube, which was heated up to 700°C, kept for 10 hours, and then slowly cooled to 450°C over 30 hours. In order to retain the β-phase, the tube was then quenched in iced water directly

from 450°C. Similarly, a series of PdBi$_{2-x}$Pb$_x$ (x= 0.0, 0.08, 0.15, 0.20, 0.28, 0.35, 0.40 0.60, 0.80, and 1.0) crystals was attempted to be grown with stoichiometric amounts of Pd, Bi, and Pb. In addition, the Na$_x$PdBi$_2$ compounds were grown by mixing Na and previously obtained PdBi$_2$ precursor, followed by the same synthetic condition but using a carbon-coated quartz tube to prevent the possible reaction of sodium with the quartz tube at high temperature. The carbon coating of the quartz tube was found to be intact after the synthesis, indicating the success of the growth. X-ray powder diffraction was performed on the powdered Pb-doped single crystals at room temperature from 10 to 90 degrees using a Panalytical X'pert Diffractometer. The actual chemical compositions were determined by wavelength-dispersive spectrometry (WDS) on a JEOL JXA-8600 electron microprobe analyzer. At least 6 well separated points across the crystal were measured to confirm the homogeneity of the dopants. The dc magnetic susceptibility was measured as a function of temperature χ(T) using a Quantum Design Magnetic Property Measurement System (MPMS) down to 2 K. Electrical resistivity as a function of temperature ρ(T) and field ρ(H) was measured by employing a standard 4-probe method using a Linear Research LR400 ac bridge operated at 15.9 Hz in a Quantum Design Physical Property Measurement System (PPMS) up to 7 T and down to 1.9 K. High pressure resistivity measurements with a 4-lead technique using the LR-400 Resistance Bridge were made in a Be-Cu pressure cell using Fluorinert FC77 as the pressure media. A lead manometer was used to measure the pressure *in situ* with the LR400 Inductance Bridge[14]. The Hall resistivity measurements were performed in the PPMS system with field up to 5T using the 5-lead technique, which balanced the longitudinal resistance to be close to zero.

**Results and discussions:**

β-PdBi$_2$ single crystals with shining metallic luster, typical size of 5mm, and preferred orientation along the *c*-axis after cleavage, could be obtained through the melt-growth technique, as shown in Fig. 1(a). The calculated lattice parameter *c* [12.963(3)Å] is consistent with previous reports[12]. A representative SEM image of the undoped β-PdBi$_2$ single crystal is also shown as the inset of Fig. 1(a). Both resistivity and magnetization measurements have shown that the superconducting transition temperature of the as-grown β-PdBi$_2$ crystal is 5.4K, indicating the improved quality of the grown crystals as also pointed out by previous reports[12]. The resistivity curve exhibits a hump below 150K and a minor downturn around 50K, as shown in Fig. 1(b), suggesting possible strong electron/spin correlation in this compound.

Fig. 2 shows the powder XRD patterns with Miller indices of the crushed crystals for the Pb-doped PdBi$_{2-x}$Pb$_x$ (x=0.08, 0.15, 0.20, 0.28 0.35, and 0.40) samples. Except for a few minor peaks of the α-phase present at low doping level, namely x=0.08, all of the peaks are well indexed into the β-PdBi$_2$-type body centered tetragonal structures in all of the samples with different doping levels. These crystals are rather stable outside of the glove box for several months. Due to the close radius sizes between the Pb and Bi, the change of lattice parameter upon Pb doping is rather small, and a gradual decrease of the lattice parameter from 12.963(3)Å for x=0 to 12.940(2)Å for x=0.4 is observed, but the overall lattice parameters change is less than 0.2%. To further confirm the homogeneity of the doped samples and establish the actual doping levels, we performed chemical analyses through WDS measurements on the single crystals. The Pb-concentration is homogenous throughout the whole samples, indicating the formation of solid solutions for these Pb-doped samples. The actual Pb-doping levels are x=0.08(2), 0.14(2), 0.19(2), 0.27(2), 0.34(1), and 0.39(1), for the nominal compositions of x = 0.08, 0.15, 0.20, 0.28, 0.35, and 0.40 in $PdBi_{2-x}Pb_x$, respectively. The results show that the actual doping levels are very close to the nominal compositions. Some extra impurity peaks belonging to the α-$PdBi_2$ monoclinic phase emerge at the doping level x=0.60 from X-ray powder diffraction and become more dominant with further Pb-doping. At the doping level of x=1.00, the XRD pattern shows the nearly pure phase of α-$PdBi_2$-type structure with no detectable β phase, implying the solubility limit of the tetragonal β-$PdBi_{2-x}Pb_x$ phase at this high doping level.

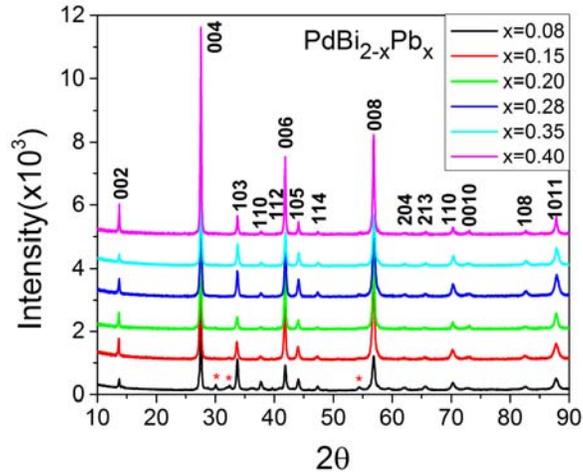

Fig.2 The powder X-ray diffraction patterns of $PdBi_{2-x}Pb_x$ with Miller indices. The patterns are vertically offset for better clarity. Some minor impurity peaks from α-phase are marked as *.

Systematic resistivity and magnetization measurements were carried out for all samples with different doping levels. As shown in Fig. 3(a), the superconducting transition temperature $T_c$ of $PdBi_{2-x}Pb_x$ continuously decreases upon Pb substitution, from 5.4K for x=0, to 4.9K for x=0.08, 4.4K for x=0.15, 3.8K for x=0.20, 2.5K for x=0.28, and 2.2K for x=0.35. The sample eventually becomes non-superconducting above 2K when the doping level reaches x=0.40. The superconducting transition width (10%–90% resistivity drop) is rather narrow (less than 0.3 K) in all samples, indicating the good quality of the doped samples.

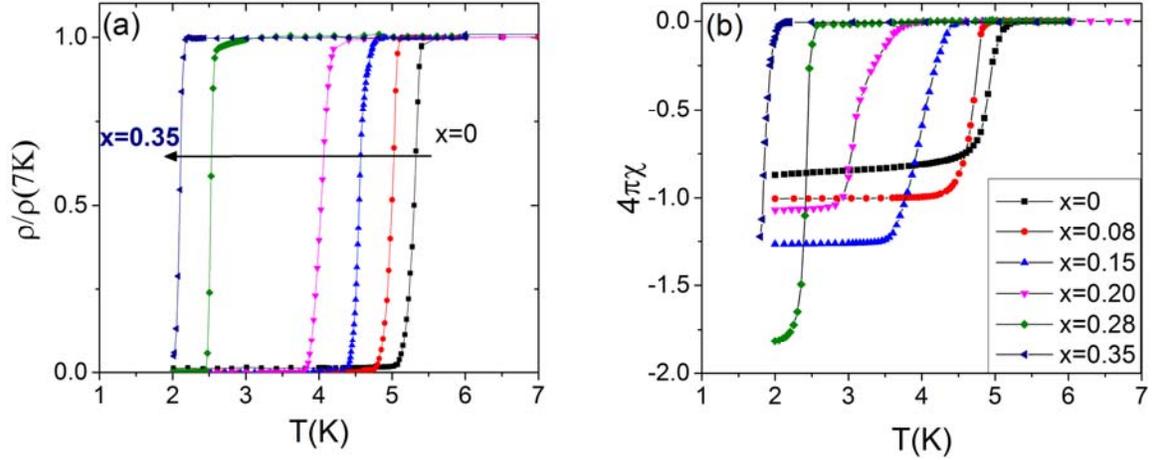

Fig.3 The resistivity (a) and magnetic susceptibility (b) data under H=2 Oe of PdBi$_{2-x}$Pb$_x$ with x=0-0.35 from 2K to 7K.

Similar consistent results were obtained through magnetic susceptibility measurements as shown in Fig. 3(b). The magnetization was measured under an applied magnetic field of 2 Oe on randomly oriented small crystals with typical mass ~20 mg packed in the gelatin capsules. All of the samples with doping level 0<x<0.35 exhibit substantial diamagnetic shifts at the lowest temperature. The shielding fractions 4πχ are close to or exceed 1 without the demagnetization factor correction, which implies the bulk superconductivity nature of these samples. Consistent with the resistivity data, the superconducting transition temperature systematically moves downward with doping from 5.2K for x=0 to 2K for x=0.40.

To confirm the effective hole doping by Pb-substitution, Hall effect measurements have been carried out to evaluate the charge carrier concentration for the parent compound PdBi$_2$ and a representative Pb-doped PdBi$_{1.8}$Pb$_{0.2}$ sample (Fig. 4). The raw Hall resistivity $\rho_H$ are linear with the field, with negative slopes for both samples. To further eliminate the effect of possible misalignment of Hall electrodes, the Hall coefficient $R_H$ was taken as $R_H = [R_H(5T) + R_H(-5T)]/2$ at each temperature. The inset of Fig. 4 shows the Hall coefficient of PdBi$_2$ from 6K to 300K. The Hall coefficient is negative over the whole temperature range and only weakly depends on temperature, suggesting that the electron-type charge carriers dominate the charge transport. The value of Hall coefficient changes by <30% from 2K to 300K, implying relatively minor multi-band effects, and therefore it is reasonable to evaluate the carrier concentration using the Hall coefficient $R_H$.

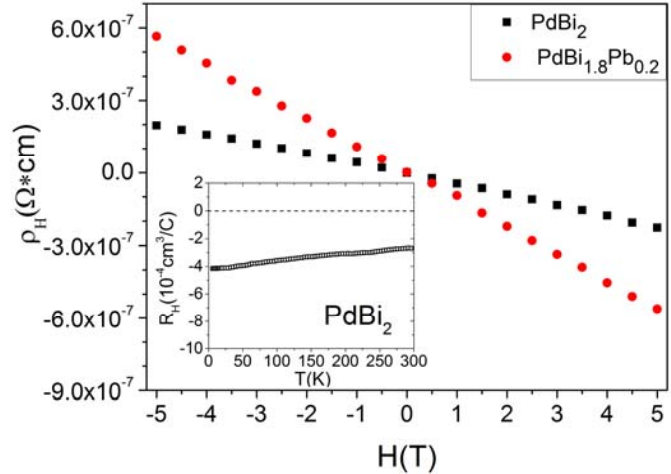

Fig.4 Hall resistivity of β-PdBi$_2$ and PdBi$_{1.8}$Pb$_{0.2}$ under magnetic field at 10K; the inset shows the Hall coefficient $R_H$ of β-PdBi$_2$ from 6K to 300K.

The Hall resistivity of the parent compound and Pb-doped sample under different magnetic fields at 10K are shown in Fig. 4, where one can see that the contribution from magneto-resistance is rather small. The Hall coefficients $R_H$, determined by the slopes of the curves, are $R_H$= -4.24*10$^{-4}$cm$^3$/C and $R_H$= -1.12*10$^{-3}$cm$^3$/C for PdBi$_2$ and PdBi$_{1.8}$Pb$_{0.2}$, respectively. The negative Hall cofficient of the Pb-doped sample indicates that the charge carriers are still dominated by electrons. By simply using the single band expression $n=\frac{1}{R_H q}$, we can calculate electron concentrations: n=1.47*10$^{22}$ cm$^{-3}$ for PdBi$_2$ and n=5.56*10$^{21}$ cm$^{-3}$ for PdBi$_{1.8}$Pb$_{0.2}$. The substantial decrease of electron concentration suggests the effective hole-doping in the PdBi$_2$ system through Pb-substitution at the Bi site. This effective hole-doping might shift the Fermi level, resulting in a lower electronic density of states (DOS). Therefore, the lower DOS might contribute to the decrease of the $T_c$ if a rigid band model is adopted.

Since the hole doping by Pb substitution caused the decrease of the $T_c$, we also attempted to carry out the Na-intercalation, which would introduce electrons into the system. Several trials with nominal Na-concentration ranging from x=0.1 to x=0.4 on Na$_x$PdBi$_2$ have been tested. X-ray powder diffraction analysis on the resulting bulk materials reveals that some small impurities (less than 10% total) of NaBi and Pd exist in the sample besides the formation of the β-tetragonal phase. We were able to isolate smaller pure crystals from the bulk samples and carried out detailed chemical analysis and physical measurements. The isolated crystals are homegeneous from WDS analysis and have all of the three elements present. However, the actual Na content is much lower than the nominal composition. Taking the nominal x=0.1 as an example, the actual composition we found from the chemical analysis was Na:Pd:Bi = 0.044(3):1:2.000(5). The highest Na doping level is ~0.057(2) determined from the chemical analysis, indicating the relatively low limit of Na intercalation into this compound compared with the Pb-doping. The Na-doped crystals were moderately sensitive to air/moisture as they slowly decayed when kept outside of the glove box for one day. This also implies the sucesssful Na-intercalation into the system. The superconducting $T_c$, on the other hand, is rapidly suppressed at such low Na-doping level, changing from 5.4K in PdBi$_2$ down to 4.1K in the Na$_{0.044}$PdBi$_2$, and further down to 3.9K in Na$_{0.057}$PdBi$_2$, as shown in the inset of Fig. 5(a). However, this is in great conflict with theoretical expectation,[13] which suggests such electron-like doping should shift the Fermi level toward the DOS peak, increasing the DOS at the Fermi level and thus the T$_c$.

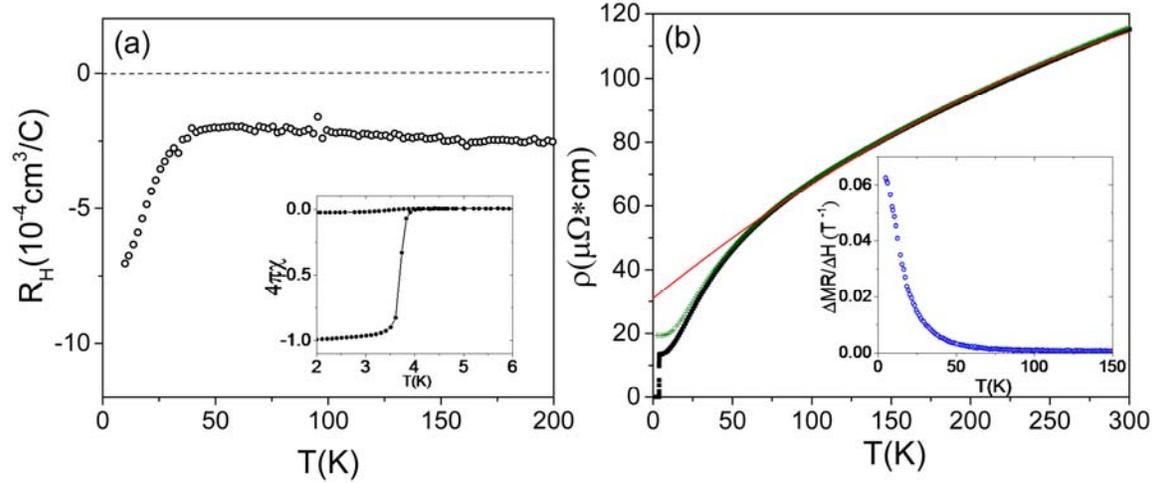

Fig.5 (a) The Hall coefficient R$_H$ of Na$_{0.057}$PdBi$_2$ from 5K to 300K; the inset shows the magnetic susceptibility of Na$_{0.057}$PdBi$_2$. (b) Magnetoresistivity of Na$_{0.057}$PdBi$_2$ from 2K to 300K (solid black squares: zero field; green crosses: 7 T); The solid red line is a quadratic function fitting of resistivity from 70K to 300K. The inset is the ratio of magnetoresistance ($\Delta$MR=[$\rho$(7T)-$\rho$(0T)]/$\rho$(0T)) divided by the magnetic field.

To verify whether the DOS does increase through Na-intercalation, the Hall measurement was carried out on the Na$_{0.057}$PdBi$_2$ sample to probe the change of charge carrier concentration. Similar to PdBi$_2$, the Hall coefficient R$_H$ above 50K is T-insensitive for the Na-intercalated sample, as shown in Fig. 5(a). Therefore, the electron concentrations for PdBi$_2$ and Na$_{0.057}$PdBi$_2$ are calculated based on the R$_H$ at 50 K as $1.58*10^{22}$ cm$^{-3}$ and $3.09*10^{22}$ cm$^{-3}$, respectively. There is apparently a signficant increase of the carrier concentration, which is also in line with the decrease of the room-temperature resistivity observed in Fig. 5(b). The deduced d$\rho$/dT at 300 K decreases from 0.37 $\mu\Omega$·cm/K for PdBi$_2$ to 0.20 $\mu\Omega$·cm/K for Na$_{0.057}$PdBi$_2$. The Na-intercalation, therefore, does introduce electrons and enhance the DOS as expected. The suppression of T$_c$ has to be attributed to other mechanisms.

It is well known that both the competing excitations and the impurity scattering (especially pair-broken scattering) may suppress superconductivity. To explore the issue, both the magnetoresistance and the R$_H$ are investigated (Fig. 5). The Hall coefficient R$_H$ is enhanced almost by a factor of three with cooling below 50 K (Fig. 5a). As demonstrated previously (Fig. 4 inset), the multiband effect has trivial interferences on the R$_H$ for the undoped PbBi$_2$. The unexpected enhancement of the R$_H$ in the Na-intercalated sample can hardly be attributed to the multiband effect, but is more likely caused by certain spin-related scattering. Spin-density-waves (SDW) under field, for example, may create local moment. This consequentially produces a Hall component similar to the anomalous Hall effects in the ferromagnetic materials but quasi-linear on H and unusually large below the SDW transition temperature. Similar situations have been reported in some pnicitides. Han *et al.* have observed that the $R_H$ of SrFeAsF significantly increased with cooling below 150-160 K with the SDW transition around 173 K[15]. Similar Hall coefficient anomalies below the SDW transition temperature have also been reported in the $AFe_2As_2$ (A=Ba, Ca, or Sr) compounds[16,17]. Therefore, the downturn of $R_H$ below 50 K tentatively suggests that a possible competing SDW transition, with a possible defect-caused broadening, emerges in the Na-intercalated $PdBi_2$. Such an interpretation seems to be supported by both the resistivity and the magnetoresistivity (Fig. 5(b)). It is well known that SDW, as an ordered ground state, is associated with an energy gap in the spin section[18]. The gap-opening below the transition temperature may appear as a conductivity increase and an enhanced magnetoresistivity associated with the gap-suppression by the magnetic field. For instance, a sharp reisistivity drop in LaOFeAs around 150 K as well as an unusually large magnetoresistivity, *i.e.* $\frac{\partial lnR}{\partial H} \approx 0.02/T$ below 50 K, have been taken as evidence for the SDW[19]. To explore the situation in $Na_{0.057}PdBi_2$, the R(T) above 70 K is fitted as the quadratic function of T (the solid red line in Fig. 5(b)), where the R-drop below 50 K is evident. The R(T) under magnetic field of 7 T, in particular, is much higher than that under 0 T (Fig. 5(b), inset). The deduced $\frac{\partial lnR}{\partial H} \approx 0.05/T$ at 10 K is comparable to that of LaOFeAs. The enhanced hump feature, the large postive magnetoresistance, and the increased amplitude of Hall coefficient below 50K suggest the presence of a spin order, possibly a SDW order, below ~ 50K. Thus the suppressed superconductivity in Na-intercalated $PdBi_2$ can be well understood as a result of the competitive SDW-like ordering. The exact nature of this induced SDW-like transtion and its interplay with the superconductivity is under investigation.

Based on the above data, we were able to construct the phase diagram of $T_c$ as a function of doping level (both hole and electron doping) as shown in Fig. 6. At the right side of the phase diagram (hole doping), the $T_c$ decreases continuously with the Pb-doping. The suppression of $T_c$ can be understood as the decrease of DOS at the Fermi level caused by hole doping. At the left side of the figure, we demonstrate that the $T_c$ is also quickly suppressed by a small amount of the Na-interclation. The cause of this $T_c$ suppression could be attributed to the emergence of the SDW-like transition, which may compete for the ground state and be counterproductive in stabilizing the superconducting state.

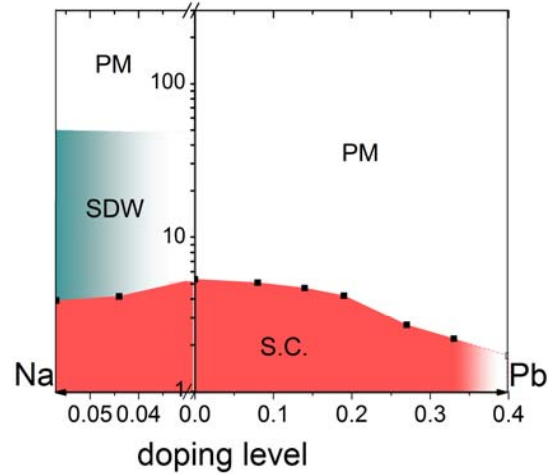

Fig.6 The phase diagram for both hole- and electron-doped PdBi$_2$. SC: superconducting transition temperature determined resistively; PM: paramagnetic state; SDW: possible spin-density-wave-like transition induced by Na-intercalation.

To further examine the above conjectures based on doping, we investigated the pressure effect on the undoped PbBi$_2$ with the highest T$_c$ of this series by measuring temperature dependence of resistivity under pressures up to 16.63 kbar, as shown in Fig. 7. As the applied pressure increases, the normal state resistivity considerably decreases. A closer look at the low temperature part, as shown in Fig.7(b), reveals that the superconducting transition gets slightly sharper, and is gradually suppressed, upon the applied pressure at a linear suppression rate of $d$T$_c$/$d$P= -0.28K/GPa. At 16.63 kbar, the T$_c$ was reduced to 4.9K. To verify the stability of the sample, the pressure cell was unloaded to lower pressure and ambient. The corresponding data are denoted as (u) in Fig. 7(b). We observed that superconducting transition T$_c$ values measured upon loading and unloading the pressure cell fell along the same line, proving the stability of the sample in the pressure cycle. The corresponding phase diagram of T$_c$ versus pressure is shown in the inset of Fig. 7(b). The suppression of T$_c$ by pressures appears to be consistent with the doping experiment.

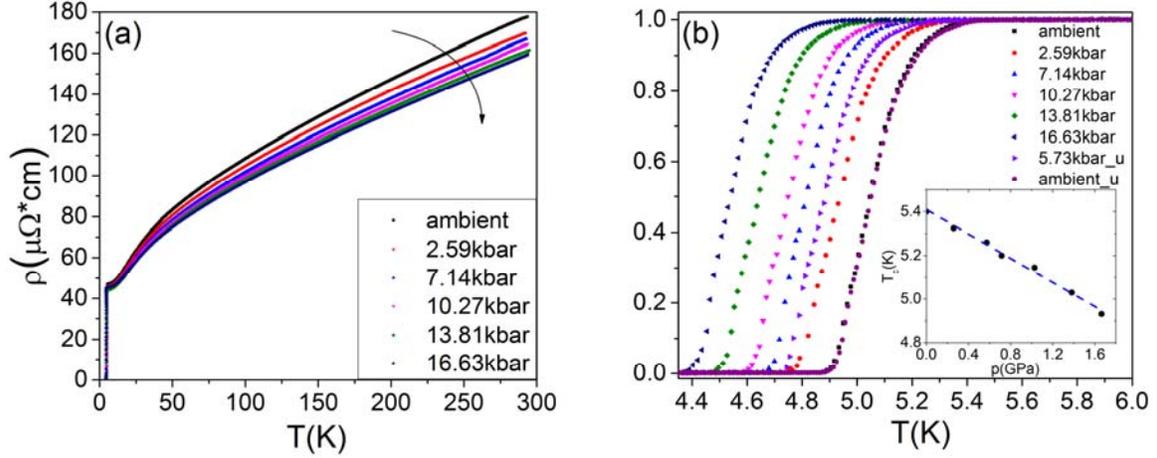

Fig.7 (a) The resistivity of β-PdBi$_2$ from 1.2K to 300K under high pressure. (b) The normalized resistivity under different pressures, where u denotes the unloaded pressure run. The inset shows the shift of T$_c$ with pressure for β-PdBi$_2$.

**Conclusions:**

In summary, we have systematically grown large single crystals of layered compound β-PdBi$_2$, the hole-doped PdBi$_{2-x}$Pb$_x$ and the electron-doped Na$_x$PdBi$_2$, and studied their magnetic and transport properties. Hall measurements on PdBi$_2$, PdBi$_{1.8}$Pb$_{0.2}$, and Na$_{0.057}$PdBi$_2$ show that the charge transport is dominated by electrons in all of the samples. The electron concentration is substantially reduced upon Pb-doping in PdBi$_{2-x}$Pb$_x$ and increased upon Na-intercalation in Na$_x$PdBi$_2$, indicating the effective hole-doping by Pb and electron-doping by Na. In Pb-doped PdBi$_2$, we observed a monotonic decrease of T$_c$ from 5.4K in undoped PdBi$_2$ to less than 2K for x > 0.35. The monotonic decrease of the T$_c$ upon doping can be explained by the reduced DOS at the Fermi level. In Na-intercalated samples, a rapid decrease of T$_c$ with a slight Na-intercalation level is also observed, which is in contradiction with the theoretical expectation. Both the magnetoresistance and Hall measurements further reveal evidence for a possible competing SDW-like transition below ~50K, which could contribute to the suppression of the T$_c$ in Na-intercalated samples. Meanwhile, application of external pressure up to 16.63 kbar on the undoped PdBi$_2$ also suppresses the superconducting transition linearly with a dT$_c$/dP coefficient of -0.28K/GPa, consistent with the doping experiments.